\documentclass[aps,prb,twocolumn,showpacs,superscriptaddress,preprintnumbers,amsmath,amssymb,longbibliography]{revtex4-1}

\usepackage{graphicx}
\usepackage{bm}
\usepackage{ulem}    
\usepackage{xcolor}
\usepackage{gensymb}                  
\usepackage{tikz}
\usepackage{mathptmx}
\usepackage{fancyhdr}
\usepackage[pdftex,colorlinks=true,pdfstartview=FitV,linkcolor=blue,citecolor=blue,urlcolor=blue]{hyperref}

\DeclareMathOperator{\e}{e}

\begin{document}

\title{Melting properties of Ag$_{x}$Pt$_{1-x}$ nanoparticles}

\author{Alexis Front}
\affiliation{Laboratoire d'Etude des Microstructures, ONERA-CNRS, UMR104, Universit\'e Paris-Saclay, BP 72, Ch\^atillon Cedex, 92322, France}

\author{Djahid Oucheriah}
\affiliation{Laboratoire d'Etude des Microstructures, ONERA-CNRS, UMR104, Universit\'e Paris-Saclay, BP 72, Ch\^atillon Cedex, 92322, France}

\author{Christine Mottet}
\affiliation{Aix-Marseille Universit\'e, CNRS, Centre Interdisciplinaire de Nanoscience de Marseille, UMR 7325, 13288 Marseille, France}

\author{Hakim Amara}
\email{hakim.amara@onera.fr}
\affiliation{Laboratoire d'Etude des Microstructures, ONERA-CNRS, UMR104, Universit\'e Paris-Saclay, BP 72, Ch\^atillon Cedex, 92322, France}
\affiliation{Universit\'e de Paris, Laboratoire Mat\'eriaux et Ph\'enom\`enes Quantiques (MPQ), F-75013, Paris, France}
%
%

\begin{abstract}

At the nanoscale, materials exhibit unique properties that differ greatly from those of the bulk state. In the case of Ag$_{x}$Pt$_{1-x}$ nanoalloys, we aimed to study the solid-liquid transition of nanoparticles  of different sizes and compositions. This system is particularly interesting since Pt has a high melting point (2041 K compare to 1035 K for Ag) which could keep the nanoparticle solid during different catalytic reactions at relatively high temperatures, such as we need in the growth of nanotubes. We performed atomic scale simulations using semi-empirical potential implemented in a Monte Carlo code at constant temperature and chemical composition in canonical ensemble. We observed that the melting temperature decreases with the size (pure systems and alloys) and the composition. We show that the melting systematically passes through an intermediate stage with a crystalline core (pure platinum or mixed PtAg depending on the composition) and a pure silver liquid skin, which strongly questions the idea of having a faceted solid particle in catalytic reactions for carbon nanotubes synthesis.

\end{abstract}


\maketitle







\section{Introduction}

The reduction of materials to the nanoscale has a profound impact on their structure and characteristics. In this context, metallic nanoparticles (NPs) are widely studied for their particular reactivity and/or rather unique physical properties in many scientific fields. Indeed, their nanometric dimension confers them different specificities compared to their bulk counterparts~\cite{Guo2013, He2018}. These specificities are enhanced in bimetallic alloys (or nano-alloys) where the association of two metals within a NP can exalt or combine certain features making their scope of application even broader~\cite{Ferrando2008, Alloyeau2012, Alloyeau2009, Amara2022}. 

A typical example is the use of bimetallic NPs as a catalyst for the synthesis of carbon nanotubes (CNTs) with a defined structure. Over the last few decades, CNTs have been the subject of intense research efforts, highlighting their exceptional mechanical, electronic, optical and thermal properties~\cite{Yang2019, Wu2021}. The main bottleneck in the development of a CNT-based technology is the difficulty of obtaining large quantities and high purity of a single CNT structure. Indeed, depending on its structure, a nanotube can be metallic or semiconducting. In terms of synthesis, they are commonly produced by catalytic Chemical Vapor Deposition (CVD) at medium temperature (800 to 1500 K) using metal NPs that catalyze the decomposition of the carbon-bearing gas precursors and make the growth of nanotubes possible~\cite{Jourdain2013, Amara2017, Rao2018}. However, the complex interaction of catalyst NPs exposed to reactive carbon makes it a challenge to achieve tube growth with a specific structure~\cite{Magnin2018, He2019, Yang2020}. For the past few years, use of bimetallic catalysts is presented by different groups as a key to achieve a structural or property control~\cite{Forel2019}. Typically, true metallic W-Co alloy NPs, seem to favour the formation of semiconducting tubes with defined chiralities~\cite{Yang2014, Yang2017}. To explain these striking results, it has been proposed that the presence of an alloying element within the catalyst with a high melting point tends to keep the particle solid during synthesis~\cite{Yang2014, Zhang2017, Li2017}. Among proposed growth mechanisms, it has been argued that the structure of the solid catalyst, such as a facets and steps, can affect the produced tubes~\cite{Zhu2005, Zhang2017}. In this approach, catalyst NPs serve as a template that could control the structure of grown carbon layers via planar or perpendicular interactions between carbon $sp^{2}$ networks and the catalyst facet, specifically relying on the epitaxial relationship. Although elegant, this explanation is open to debate as there are only few experimental studies to determine the state of the catalyst during synthesis. Experimental investigations at this level are usually done by Transmission Electron Microscopy (TEM) performed \textit{post mortem} outside the CVD reactor~\cite{Zhang2017, An2019, Zhang2022, Li2022}. Besides, \textit{in situ} TEM reveals more details about the role of the catalysts during the CNT nucleation-growth~\cite{Hofmann2007, Picher2014, Zhang2017b}. However, direct evidence of the state of the NP using this approach is lacking and this even if a perfectly crystalline particle seems to be highlighted very recently during the CNT nucleation and thus appearing to confirm the model proposed so far. Although these techniques are very advanced, the precise state of the surface (solid, liquid or a mixture) is not so easy to identify and are very scarce. This is particularly true in the typical temperature range of CNT synthesis. In order to address this issue, we have chosen to study the Ag-Pt system, which is particularly interesting in the field of catalysis in general~\cite{Esfandiari2016}. Although this alloy is not used for carbon nanotube growth, it is an ideal model system since it involves two metals with very different melting temperatures, i.e. 1235 K and 2041 K for Ag and Pt, respectively. 

The aim of the present work is to investigate at the atomic scale the state of Ag$_{x}$Pt$_{1-x}$ nanoparticles (solid, liquid, mixed?) of varying sizes and compositions during the different stages of their melting. For this purpose, we performed atomic scale simulations using a semi-empirical potential implemented in a Monte Carlo code to lead the structure towards its thermodynamic equilibrium~\cite{Frenkel2002} and to analyse in details the solid-liquid transitions. Obviously, there are many works in the literature seeking to characterize the state of the particle during the solid-liquid transition for pure metals~\cite{Delogu2005, Omid2011, Fu2017, Samantaray2021} or alloys~\cite{Kim2007, Omar2016}. Indeed, Buffat and Borel revealed in the 1970s a significant decrease in the melting temperature of very small gold NPs, explaining that the surface-to-volume ratio was at the origin of this phenomenon~\cite{Buffat1976}. Regarding the case of bimetallic NPs, the study of systems with two different melting points is badly lacking, not allowing to show if the crystalline structure of the nanoparticle is preserved at relatively high temperature by the combination of these two metals within the NP. Moreover, the analysis of the surface during the solid-liquid transition has never been addressed. Based on our atomic scale study, we confirmed that the melting temperature decreases with the size of the NPs (pure systems and alloys). However, our detailed analysis shows that the melting systematically passes through an intermediate stage with a crystalline core (pure Pt or AgPt depending on the composition) and a layer of liquid Ag, which strongly questions the idea of having a faceted solid particle in catalytic reactions at high temperature.

\section{Methodology}

\subsection{Ag-Pt system}
\label{sec:sec1}

The bulk Ag-Pt system presents an atypical phase diagram in the Ag-based family, with a large miscibility gap between Ag$_{2}$Pt$_{98}$ and Ag$_{88}$Pt$_{12}$ at low temperature~\cite{Okamoto, Durussel} and a L1$_{1}$ phase for an extremely narrow composition range around Ag$_{50}$Pt$_{50}$~\cite{Hart}. This phase is quite scarce in the face-centered cubic alloys with an alternation of pure planes in the (111) direction instead of the more usual L1$_{0}$ phase. 

\begin{figure}[h]
\centering
  \includegraphics[height=2.8cm]{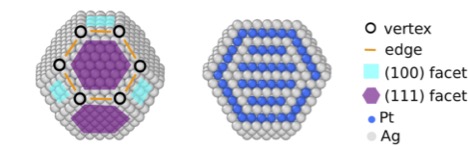}
  \caption{Core@shell L1$_{1}$/Ag/Pt@Ag structure of a Truncated Octahedron (TOh) nanoalloy of 1289 atoms. The Ag-shell (left) is composed of vertex, edge, (100) and (111) facets. The core (right) is characterized by a L1$_{1}$ phase surrounded by a subsubsurface enriched in Ag and a subsurface enriched in Pt, named L1$_{1}$/Ag/Pt.}
  \label{fgr:fig0}
\end{figure}

At the nanoscale, Ag-Pt system has been recently investigated experimentally by High-Resolution TEM. In case of Ag$_{3}$Pt, Truncated Octahedron  (TOh) NPs up to 3 nm exhibit a L1$_{1}$ ordered phase in the core and a silver surface shell whereas for larger ones the ordered phase breaks in multiple domains or is interrupted by faults~\cite{Pirart}. With increasing particle size, the Ag-shell causes internal stress in the L1$_{1}$ domains leading to subsurface enrichment in Pt coupled to a "subsubsurface" enrichment in Ag, that allows to release the inhomogeneous stress~\cite{Front}. As a result, the core@shell L1$_{1}$/Ag/Pt@Ag structure (Fig. \ref{fgr:fig0}) is therefore stabilized for size bigger than 3 nm or 807 atoms. In the present work, pure Ag and Pt NPs as well as Ag$_{3}$Pt, AgPt and AgPt$_{3}$ NPs will be considered to characterize the effect of Pt concentration on the melting temperature of NPs of different sizes (from 201 atoms to 2951 atoms).

\subsection{Tight-Binding potential}

The energetic model is based on a semi-empirical many-body potential derived from the density of states of the $d$-band metals approximated to its second moment (width of the $d$-band) in the tight-binding framework~\cite{Ducastelle1970, Rosato} (TB-SMA). It leads to an attractive term with a square root dependence on the coordination number for the band term at site $i$: 

\begin{equation}
E^{\mathrm{band}}_{i} =-\sqrt{\sum_{j,r_{ij}<r_{ab}^{\mathrm{cut}}}\xi_{ab}^{2}\e^{-2q_{ab}\left(\frac{r_{ij}}{r_{ab}^{0}}-1\right)}}
\end{equation}

where $(a,b)$ is the nature of the metal atom, $r_{ij}$ is the distance between the atom at site $i$ and one neighbor at site $j$, $r_{ab}^{\mathrm{cut}}$ is the cutoff distance, and $r_{ab}^{0}$ is the first-neighbor distance depending on the nature of the atoms. An empirically repulsive term of the Born-Mayer type is added to simulate the ionic and electronic repulsions: 

\begin{equation}
E^{\mathrm{rep}}_{i} =\sum_{j,r_{ij}<r_{ab}^{\mathrm{cut}}}A_{ab}\e^{-p_{ab}\left(\frac{r_{ij}}{r_{ab}^{0}}-1\right)}
\end{equation}

Ag-Pt system has the particularity to form an ordered L1$_{1}$ phase at the equiconcentration which is stabilized by taking into account the contribution of the second-neighbors~\cite{Front}. Consequently, the TB-SMA potential has been adapted by adding an attractive term with a gaussian shape centered on the second-neighbor distance $r^{2nd}_{ab}$: 
\begin{equation}
E^{\mathrm{gauss}}_{i} = -\sum_{j,r_{ij}<r^{\mathrm{cut}}_{ab}\atop a\neq b}G_{ab}\e^{-\frac{(r_{ij}-r^{2nd}_{ab})^{2}}{2{\sigma_{ab}}^{2}}}
\end{equation}

Therefore, the total energy of an atom $i$ is given by:

\begin{equation}
E^{\mathrm{tot}}_{i} = E^{\mathrm{band}}_{i} + E^{\mathrm{gauss}}_{i} + E^{\mathrm{rep}}_{i}
\end{equation}

The homo-atomic ($a=b$) parameters are fitted to the cohesive energies, lattice parameters and elastic constants calculated by density functional theory (DFT). For the hetero-atomic interactions ($a\neq b$), the parameters $p_{ab}$ and $q_{ab}$ are the average of the pure metal ones. Finally, $A_{ab}$, $\xi_{ab}$, $G_{ab}$ and $\sigma_{ab}$ are fitted to the formation enthalpies of the L1$_{0}$ and the L1$_{1}$ phases as well as the solution energies of one impurity in the matrix of the other metal. More details on the fitting procedure can be found in Ref.~\cite{Front}

\begin{table}[h]
\small
  \caption{\ Parameters of the Tight-Binding SMA potential.}
  \label{tbl:example1}
  \begin{tabular*}{0.48\textwidth}{@{\extracolsep{\fill}}llllllll}
    \hline
    a-b & $p_{ab}$ & $q_{ab}$ & $A_{ab}$ (eV) & $\xi_{ab}$ (eV) & $G_{ab}$ (eV) & $\sigma_{ab}$ &  \\
    \hline
    Pt-Pt & 10.7960 & 3.1976 & 0.1993 & 2.2318 & - & - & \\
    Ag-Ag & 11.7240 & 2.8040 & 0.0748 & 1.0064 & - & - & \\
    Pt-Ag & 11.2600 & 3.0008 & 0.1456 & 1.5920 & 0.0440 & 0.1393 & \\
    \hline
  \end{tabular*}
\end{table}

\subsection{Monte Carlo simulations}

This atomic interaction model is then implemented in a Monte Carlo (MC) code, based on the Metropolis algorithm \cite{Metropolis} using the canonical ensemble. This procedure makes it possible to relax the structure at finite temperature according to a Boltzmann type probability distribution. In the canonical ensemble, standard MC trials correspond to randomly exchange the positions of two atoms with a different nature or randomly choosing an atom and its displacement with an amplitude $0.05\sqrt T\left(\xi-0.5\right)$. $T$ is the temperature and $\xi$ is a random number between 0 and 1. We performed 10$^{3}$ MC macrosteps for equilibration then the average quantities are calculated over 10$^{3}$ macrosteps. One macrostep consists in proposing randomly either $n_{\mathrm{dep}}*N$ displacements of atom or $n_{\mathrm{ex}}*N$ atomic exchanges, $N$ being the total number of atoms of the system. In practice, we chose $n_{\mathrm{dep}}=10$ and $n_{\mathrm{ex}}=1$ to ensure a quick convergence of the total energy. This quantity is taken as a sum of local terms, this avoids recalculating the total energy of the whole system at each Monte Carlo step making efficient the implementation of our TB-SMA model. As a result, the total energy is only recalculated at each MC trial for atoms impacted by the displacement or the exchange of an atom $i$ as it is the case here. This approach is then perfectly adapted to deal with large systems. We performed heating (increasing temperature) Monte Carlo simulations which means the simulation at the next temperature starts from the last converged configuration of the previous temperature.

\section{Melting of pure nanoparticles}

\begin{figure}[h]
\centering
  \includegraphics[height=12cm]{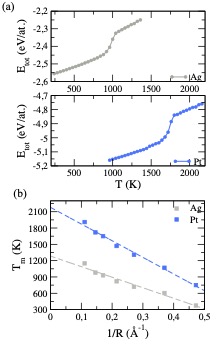}
  \caption{(a) Caloric curves of pure Ag and Pt NPs. (b) Melting temperature of pure Ag (in grey) and pure Pt (in blue) nanoparticles of TOh morphologies as a function of the inverse radius.}
  \label{fgr:fig1}
\end{figure}

First, the finite size effect of pure Ag and Pt NPs on the melting temperature ($T_{\mathrm{m}}$) is investigated. In this context, we considered TOh shapes from 201 to 2951 atoms corresponding to magic numbers. The solid-liquid transition is characterized by performing simulated annealing where the typical trend of total energy as a function of temperature (see Fig.~\ref{fgr:fig1}a). Starting from a crystalline structure, these caloric curves display an initial rapid increase in energy with temperature. At the melting point, a large jump is observed indicating the phase change corresponding to the melting of the NPs. Then, a steady increase in energy with temperature is observed where the NP is liquid. Most of the time, in the past, these curves have been obtained by molecular dynamic simulations~\cite{Ercolessi91,Lewis97,Goddard01,Delogu05,Mottet05} and further analyzed in terms of phase coexistence in finite systems~\cite{Wales94,Hendy05} to be compared to first-order melting transition in bulk systems. Here the same behavior is obtained by canonical Monte Carlo simulations and clearly indicates a two-state model (solid/liquid) at the phase transition where $T_{\mathrm{m}}$ can be obtained.

Using this approach, we determined $T_{\mathrm{m}}$ for Ag and Pt clusters with different size (from 201 to 2951 atoms) in the truncated octahedral shape. $T_{\mathrm{m}}$ is plotted against $1/R$ (Fig.~\ref{fgr:fig1}b) where $R$ is the radius of the nanoparticle defined as $R=\left(N/4\right)^{\frac{1}{3}}$, with $N$ the number of atoms. The melting temperature exhibits a linear decrease  as a function of $1/R$, in agreement with the first experimental observation by Buffat and Borel~\cite{Buffat1976} of gold nanoparticles in a size range of 2 to 20 nm in diameter. Later on, Frenken and van der Veen~\cite{Frenken85} observed the melting on a (110) lead surface at lower temperature than the lead melting point, called also "surface premelting", which can be correlated to the liquid phase formed at the surface of nanoparticles. It has been accepted that beyond a certain size, where surface and core can be distinguished, the formation of a liquid surface skin was a precursor effect of the melting of NPs.~\cite{Ercolessi91,Goddard01,Delogu05} Therefore the surface/volume ratio variation with cluster size determines the decrease of melting temperature with cluster size. 

 From the two curves of melting temperature as a function of cluster size in Fig.~\ref{fgr:fig1} we can extrapolate the melting temperature of the bulk, leading to 1280~K for Ag and 2170~K for Pt, in excellent agreement with the experimental values, 1235~K and 2041~K respectively. These results show that the interatomic potential fitted on bulk data is perfectly transferable to nanoparticles and thus well adapted to our problem where we seek to characterize the solid-liquid transitions of nano-alloys involving elements with two very different melting points.  

\subsection{TOh$_{1289}$ AgPt$_{3}$}

We now focus on bimetallic AgPt$_{3}$ NPs and analyse chemical arrangement evolution as a function of the temperature by performing simulated annealing on a truncated octahedron (TOh) nanoalloy containing 1289 atoms with a composition of 25$\%$ of Ag. 

\begin{figure}[h]
\centering
  \includegraphics[height=9.2cm]{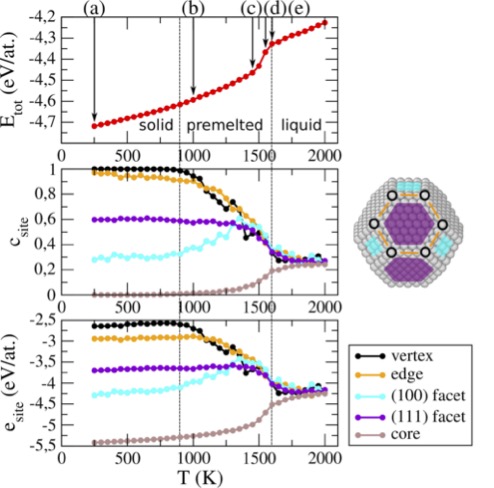}
  \caption{Total energy (top), site concentration of Ag (middle) and site energy (bottom) as a function of the temperature of truncated octahedron of 1289 atoms at AgPt$_{3}$ composition. The structures corresponding to the temperatures pointed by (a-e) are represented in Fig. \ref{fgr:fig3}. Solid, premelted and liquid zones are separated by dashes lines.}
  \label{fgr:fig2}
\end{figure}

\begin{figure}[h]
\centering
  \includegraphics[height=10cm]{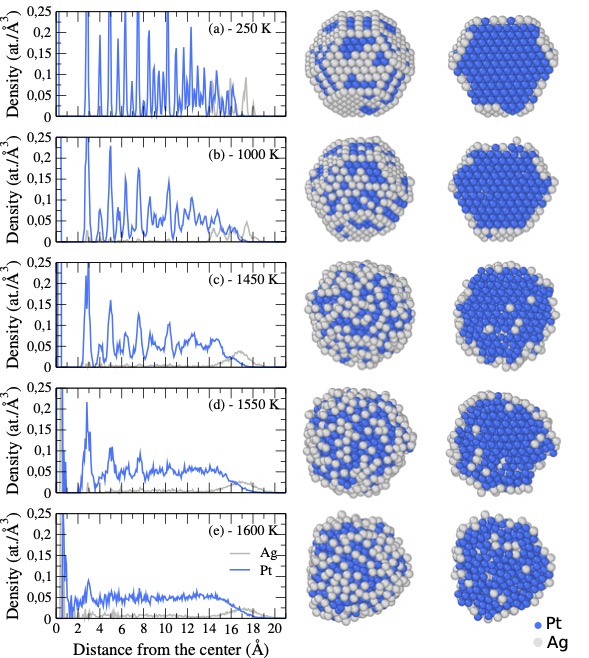}
 \caption{Atomic density of Ag (in grey) and of Pt (in blue) as a function of the distance from the center of TOh$_{1289}$ nanoparticles of AgPt$_{3}$ composition at 250 K (a), 1000 K (b), 1450 K (c), 1550 K (d) and 1600 K (e) as reported in Fig. \ref{fgr:fig2}. The middle column represent shell views, and the right column shows cutting views of the middle column. Ag atoms are in grey and Pt atoms in blue.}
  \label{fgr:fig3}
\end{figure}

As seen in Fig.~\ref{fgr:fig2}, three different regimes are observed. The first zone is characterized by a gradual increase of the total energy with temperature where the NP is in the solid state. From $\sim$ 900~K, a change of slope is observed until a jump feature of the melting temperature (here around 1500~K). Lastly, the NP is liquid and a regular increase of the total energy is again recorded. To go beyond, local analyses are performed by addressing the concentration and energy of the different sites during the MC simulations. From the site concentrations curves we notice that the initial configuration exhibits a Pt pure core and a mixed Ag-Pt surface configuration with Ag pure edges and vertices, and mixed Pt-Ag (111) and (100) facets with an Ag composition of 0.60 and 0.33 respectively.~\cite{Front22,FrontPhD}. It is worth to be noticed that usually the surface segregation (here Ag segregation) is larger for less coordinated sites. The exceptional inversion of hierarchy between the (111) and (100) facets is due to the chemical ordering on the (100) facets which forms atomic rows of alternatively Ag/Pt. Because of the size of the facet, three by three atoms square, and because the edges are populated by Ag atoms, we get two atomic rows in Pt and one in Ag (see Fig.~\ref{fgr:fig3}a) leading to the composition of 1/3 in Ag. This composition is lower than the one of the (111) facets which are fully disordered.~\cite{Front22,FrontPhD} In the solid regime, the concentrations remain constant and decrease up to the nominal concentration (25$\%$ of Ag) starting with the vertices, the edges and the facets. However the two types of facet do not evolve in the same way. Whereas the (111) facet concentration decreases regularly, after the vertices and edges, the (100) facet concentration, first increases to reach the same value as the other ones, and then decreases with the other ones in order to get an homogeneous Ag concentration on the whole surface around 1300~K. Also at that temperature the core begins to incorporate Ag atoms and at the end of the melting process, in the liquid phase, the core concentration is just slightly lower that the surface site one.  

The local energies (still in Fig.~\ref{fgr:fig2}) follows globally the concentration behaviors. The lowest energies around -5.50 eV/at. correspond to the core Pt atoms which have 12 neighbors. This is very near the total energy per atom of a face-centered cubic bulk Pt in our TB-SMA model (-5.53 eV/at.~\cite{Front}). The second population corresponds to the surface atoms with higher energy. It is noteworthy that the site energy hierarchy according to the coordination number ($Z_{i}$) is partially respected because of chemical composition. In a pure system, the larger $Z_{i}$ is, the higher the cohesion energy leading to the following hierarchy: $e_{\mathrm{(111)}}<e_{\mathrm{(100)}}<e_{\mathrm{edge}}<e_{\mathrm{vertex}}$. As already mentioned, because of the surface chemical ordering on the (100) facet with alternative Ag/Pt rows, the (100) facets are 2/3-rich in Pt, which makes it more stable by the presence of a high concentration of Pt whose cohesion energy is lower leading to a lower energy for the (100) facets as compared to the (111).  When the temperature increases, going through the premelted regime, we can see the destabilization of the most sub-coordinated atoms leaving their equilibrium positions. Indeed, the site energies of the edges and vertices decrease strongly contrary to the more coordinated sites as core and facet ones which are less affected. In the liquid regime, vertices, edges, and facets no longer exist leading to an energy distribution relatively homogeneous and an average Ag concentration equal to the nominal one.  Consequently, this local analysis shows us that the intermediate zone between 900~K to 1600~K corresponds to a destabilization of the surface which thus becomes less crystalline rather quickly before the complete melting of the NP.

The analysis of the density profiles along the radius of the NP displayed in Fig.~\ref{fgr:fig3} confirms in a more quantitative way such observation and helps to move a step further. At low temperature, the solid NP is characterized by an alternance of peaks with a full Pt core and above 14~\AA \space a mixed Ag-Pt surface. This chemical arrangement is clearly illustrated in Fig.~\ref{fgr:fig3} where snapshots of the TOh$_{1289}$ AgPt$_{3}$ NPs and its slice view are presented. At 1000~K, the chemical configuration (Pt core and Ag-Pt surface) is kept with slightly wider peaks due to the higher temperature. More precisely, the peaks are broadening near the surface, whereas they remains narrow in the core.  Notably, the Ag peak at the surface is quite large which is the signature of a less pronounced crystallinity as seen on the slice view where the surface is strongly distorted. When increasing $T$, no more sharp peaks (Ag and also Pt) are present mainly between 14~\AA \space and 19~\AA \space corresponding to a melted surface. At higher temperature ($T=$ 1550~K), the melting process diffuses within the NP where only the center of the NPs (to 8~\AA) is still solid whereas the remaining NP is completely liquid. Finally, a complete melting of the NP is achieved at $T>T_{\mathrm{m}}$ where no peak is observed anymore but Ag surface segregation remains as in the solid state. 

Thanks to our very detailed analysis, we have shown that the solid-liquid transition goes through a premelted phase of the surface which propagates along the NP.  In the present case, the complete melting of the nanoparticle is estimated at a temperature of 1600~K. Interestingly, at much lower temperature (here around 1300~K), the surface is completely distorted so that no facets are present anymore. Note that similar conclusions were obtained whatever the size of the NP. 

\subsection{TOh$_{1289}$ Ag$_{3}$Pt}

\begin{figure}[h]
\centering
  \includegraphics[height=9.2cm]{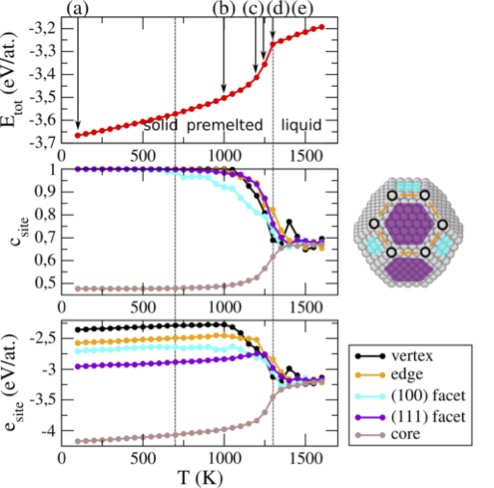}
  \caption{Total energy (top), site concentration of Ag (middle) and site energy (bottom) as a function of the temperature of TOh$_{1289}$ at the composition Ag$_{3}$Pt. The structures corresponding to the temperatures pointed by (a-e) are represented in Fig. \ref{fgr:fig5}. Solid, premelted and liquid zones are separated by dashes lines.}
  \label{fgr:fig4}
\end{figure}

\begin{figure}[h]
\centering
  \includegraphics[height=10cm]{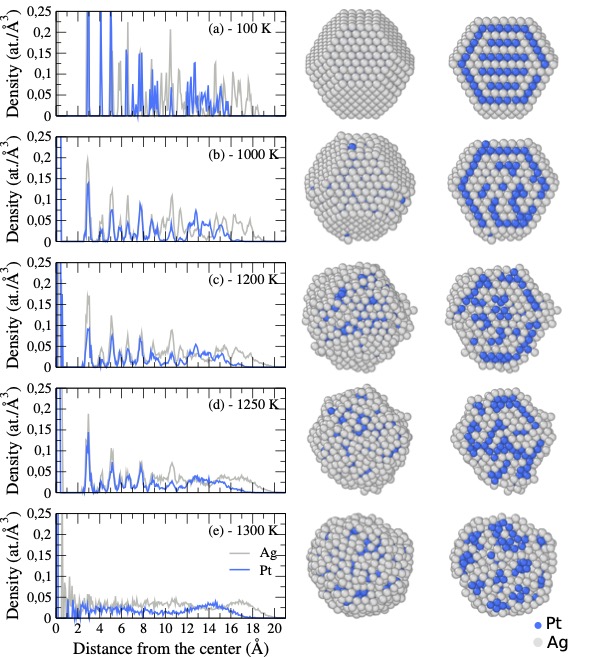}
  \caption{Atomic density of Ag (in grey) and of Pt (in blue) as a function of the distance from the center of TOh$_{1289}$ nanoparticles of Ag$_{3}$Pt composition at 100 K (a), 1000 K (b), 1200 K (c), 1250 K (d) and 1300 K (e) as reported in Fig. \ref{fgr:fig4}. The middle column represent shell views, and the right column shows cutting views of the middle column. Ag atoms are in grey and Pt atoms in blue.}
  \label{fgr:fig5}
\end{figure}

We extend the previous analysis to the Ag$_{3}$Pt NPs where all results for TOh$_{1289}$ are represented in Fig.~\ref{fgr:fig4} and Fig.~\ref{fgr:fig5}. The initial configuration at 100~K is a core@shell L1$_{1}$/Ag/Pt@Ag as described in section \ref{sec:sec1} and illustrated in Fig.~\ref{fgr:fig5}-a. 

When increasing the temperature, we again found the three regions discussed before, namely the solid, premelted and liquid ones. Compared to the AgPt$_{3}$ stoichiometry, the premelted threshold is lowered: $\sim$700~K vs $\sim$900~K. This is mainly due to the absence of Pt atoms on the surface with higher melting point. Consequently the melting temperature is also strongly impacted with a decrease of about 300~K. Regarding the local analysis, the hierarchy of the site is well respected in that case since the shell is fully composed of Ag atoms. However, during the simulation annealing, in the premelted zone, (100) facets are first impacted with a slight Pt enrichment (see ~Fig.\ref{fgr:fig4}). Then the low-coordinated atoms (vertices and edges)  and the (111) facets leave their equilibrium positions strongly modifying the structure of the surface. As found in case of AgPt$_{3}$ NP, we get a completely liquid state above $T_{\mathrm{m}}$ where the notion of site is no longer relevant with an average Ag concentration equal to the nominal one, i.e. of about 0.75. To go further, the atomic density profiles of Ag and Pt along the radius of the NP at different temperatures are presented in Fig.~\ref{fgr:fig5}. The initial configuration reveals quite sharp peaks as expected from a crystalline structure. More precisely, the presence of a pure Ag surface is perfectly clear with alternating Ag and Pt peaks typical of the core@shell L1$_{1}$/Ag/Pt@Ag structure detailed earlier. At 1000~K, the surface is no longer Ag pure with a slight Pt enrichment on the (100) facets seen in Fig.~\ref{fgr:fig4}. The slice views highlight a disordered core,  keeping a subsurface enriched in Pt and a subsubsurface enriched in Ag. When increasing the temperature ($T=$~1200~K), atomic density profiles display broad peaks in the surface area (between 16~\AA \space and 19~\AA). This particular state corresponds to a surface shell containing (111) facets partially melted coupled to other sites fully melted. Meanwhile, the premelted surface propagates within the NP through the (100) facets leading to disordered layers beneath.
At 1250 K, the core of the NP is fully cristalline whereas no more peaks are observed between 11~\AA \space and 20~\AA \space, signature of a melted surface and subsurface. When the solid-liquid transition is achieved, at 1300~K, all peaks have disappeared from the atomic density profiles revealing a homogeneous structure along the particle. More precisely, the liquid state is composed of an Ag enriched surface with Pt atoms underneath. Thus, it seems that the chemical arrangement observed in the liquid state with Ag segregation and Pt subsurface segregation remains  from the initial crystalline structure.

As highlighted in case of AgPt$_{3}$ NPs, the same melting mechanisms are observed with first of all a premelted surface which propagates within the NP. Here again, the presence of Pt obviously increases the $T_{\mathrm{m}}$ of the NP as compared to Ag pure ones. However, the melting of the surface is observed far before the complete melting  leading to a facet loss at relatively low temperature (a few hundreds of Kelvin below the melting temperature) and this whatever the size of the NP considered. 

\subsection{Generalization to the Ag$_{x}$Pt$_{1-x}$ system}

\begin{figure}[h]
\centering
  \includegraphics[height=5.3cm]{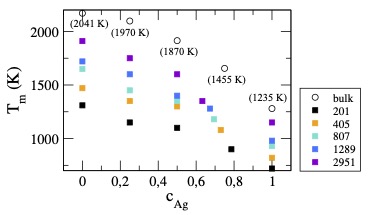}
  \caption{Melting temperature as a function of size and composition of TOh Ag$_{x}$Pt$_{1-x}$ NPs and their bulk counterparts. In case of bulk structures, the experimental data~\cite{Okamoto} are given in bracket.}
  \label{fgr:fig6}
\end{figure}

In the following, we extend the melting temperature determination to nanoparticles with sizes from 201 atoms to 2951 atoms and also of equiatomic composition AgPt. In this particular case, the TOh structure is characterized by a Ag shell and a Pt-rich core containing (111) Ag-planes~\cite{Front}. The calculated $T_{\mathrm{m}}$ for different compositions and sizes, including bulk phases, are presented in Fig. \ref{fgr:fig6}. 
 
First of all, we can note that our TB-SMA model correctly reproduces the melting temperature of the bulk phases which increases with the platinum concentration. This good agreement makes relevant our approach to characterize the $T_{\mathrm{m}}$ of NPs at different compositions. At a given concentration, the melting temperature decreases with size as observed in case of pure Ag and Pt. Typically, the melting temperature is almost divided by a factor two from bulk phase to TOh$_{201}$ NPs. For the largest sizes (here 2951 atoms), $T_{\mathrm{m}}$ is close to that of the bulk whatever the concentration. Moreover, the detailed analyses presented before have clearly shown that the melting starts from the surface and then propagates within the NP. As for pure metals, the surface/volume ratio drives $T_{\mathrm{m}}$ which therefore decreases with the size. Meanwhile, the melting temperature clearly increases with Pt concentration at a given size. However, our calculations emphasize that the variation is not strictly linear. Indeed, two zones are perfectly defined, with a frontier at the equiconcentration. On the Ag-rich side, $T_{\mathrm{m}}$ shifts significantly towards higher temperatures when the Pt concentration is raised. Indeed, it increases on average by about 30\% when moving from pure Ag to Ag$_{3}$ Pt NPs. However, increasing the amount of Pt atoms on the Pt-rich side has less impact on the melting temperature. Thus we observe an improvement of less than 10\% from AgPt$_{3}$ to pure Pt.  It is worth noticing that $c_{\mathrm{Ag}}$ varies as a function of cluster size in order to fit the concentration of the L1$_{1}$@Ag phase.

Obviously, the presence of Pt in the NP enhances the melting temperature, but the effect is more sensible in the Ag-rich side than in the Pt-rich side, so it is not necessary to consider alloys with a very high Pt concentration to delay the solid-liquid transition. 

\section{Conclusions}

Through this study on the solid-liquid transition of Ag$_{x}$Pt$_{1-x}$ NPs, several significant insights have been highlighted that may be of interest to the research community dealing with the control of  nanotube structure grown from metal particles and in the field of nanoalloys in a more general way. Our atomic scale simulations clearly show that the presence of Pt, which has a high melting point, results in a higher solid-liquid transition than in the case of a pure Ag NP. This conclusion is therefore in direct line with the experimental interpretations put forward to justify the use of bimetallic particles with a high melting point element to grow tubes on NPs within the solid state. Nevertheless, our very detailed study has also provided more precise information on the structure of the nanoparticle before its complete melting. Indeed, some complex transitions have been revealed from the surface to the core of the nanoparticle whatever the size and the composition. More precisely, we systematically observed a liquid Ag layer forming at the surface at temperatures well below the complete solid-liquid transition whatever the size and the composition of the NP. In other words, although Pt tends to delay the melting point of the NP, it cannot prevent the presence of a liquid surface where the rest of the particle remains crystalline. In such a core(solid)/shell(liquid) structure, the facets or the steps are no longer present in contradiction with the idea of a possible epitaxial growth of CNTs.   

Although Ag$_{x}$Pt$_{1-x}$ system was clearly not considered for the growth of carbon nanotubes, our study shows that the assumptions proposed by the community to explain the excellent tube selectivity from bimetallic NPs is not so simple. Obviously, these conclusions are very dependent on the system studied here. Typically, the strong tendency of the element with lower melting point (here Ag) to segregate necessarily influences the mechanisms of melting by the surface. As a result, it is not the purpose of the present work to generalize our results to the different bimetallic alloys used in the CVD growth of carbon nanotubes. Each catalyst being unique, a complete and precise study of its physico-chemical state combining both atomic scale simulation~\cite{Qiu2019} and experimental analysis is therefore essential to engineer nanocatalyst with defined properties.

\section*{Acknowledgements}

The authors thank ANR GiANT (N$\degree$ANR-18-CE09-0014-04). The authors acknowledge networking support from the International Research Network "Nanoalloys" of CNRS. This work was granted access to the HPC resources of IDRIS under the allocation 2017-096829 made by GENCI. H.A. thank C. Bichara for fruitful discussions.

\end{document}